\documentclass[twocolumn,showpacs,preprintnumbers,amsmath,amssymb]{revtex4}
\usepackage{graphicx}
\usepackage{dcolumn}
\usepackage{bm}
\begin{document}


\title{Searching for cosmic missing baryons 
with DIOS \\ 
-- Diffuse Intergalactic Oxygen Surveyor --}

\author{Yasushi Suto\footnote{suto@phys.s.u-tokyo.ac.jp}, 
 Kohji Yoshikawa\footnote{kohji@utap.phys.s.u-tokyo.ac.jp}}
\affiliation{%
Department of Physics, School of Science, University of
Tokyo, Tokyo 113-0033, Japan
}%

\author{Noriko Y. Yamasaki\footnote{yamasaki@astro.isas.jaxa.jp},
Kazuhisa Mitsuda\footnote{mitsuda@astro.isas.jaxa.jp}, 
Ryuichi Fujimoto\footnote{fujimoto@astro.isas.jaxa.jp}, 
Tae Furusho\footnote{furusho@astro.isas.jaxa.jp}}
\affiliation{
Institute of Space and Astronautical Science (ISAS), Japan Aerospace
Exploration Agency (JAXA), 3-1-1 Yoshinodai, Sagamihara, Kanagawa 229-8510,
Japan
}%

\author{Takaya Ohashi\footnote{ohashi@phys.metro-u.ac.jp}, 
Manabu Ishida\footnote{ishida@phys.metro-u.ac.jp},
Shin Sasaki\footnote{sasaki@phys.metro-u.ac.jp}, 
Yoshitaka Ishisaki\footnote{ishisaki@phys.metro-u.ac.jp}
}
\affiliation{
Department of Physics, Tokyo Metropolitan University,
1-1 Minami-Osawa, Hachioji, Tokyo 192-0397, Japan
}%

\author{Yuzuru Tawara\footnote{tawara@u.phys.nagoya-u.ac.jp}, 
Akihiro Furuzawa\footnote{furuzawa@u.phys.nagoya-u.ac.jp}}
\affiliation{
Department of Physics, Nagoya University, 
Furo-cho, Chikusa-ku, Nagoya 464-8602, Japan
}%

\date{\today}%

\begin{abstract}
Approximately 30 to 50 percent of the total baryons in the present
universe is supposed to take a form of warm/hot intergalactic medium
(WHIM) whose X-ray continuum emission is very weak.  In order to carry
out a direct and homogeneous survey of elusive cosmic missing baryons,
we propose a dedicated soft-X-ray mission, {\it DIOS} (Diffuse
Intergalactic Oxygen Surveyor). The unprecedented energy resolution
($\sim 2$eV) of the XSA (X-ray Spectrometer Array) on-board {\it DIOS}
enables us to identify WHIM with gas temperature $T=10^{6-7}$K and
overdensity $\delta=10-100$ located at $z<0.3$ through emission lines of
O{\sc vii} and O{\sc viii}.  {\it DIOS}, hopefully launched in several
years, promises to open a new window of detection and characterization
of cosmic missing baryons, and to provide yet another important and
complementary tool to trace the large-scale structure of the universe.
\end{abstract}

\maketitle

\section{Introduction}

\begin{figure*}[thb]
\begin{center}
\includegraphics[width=14cm]{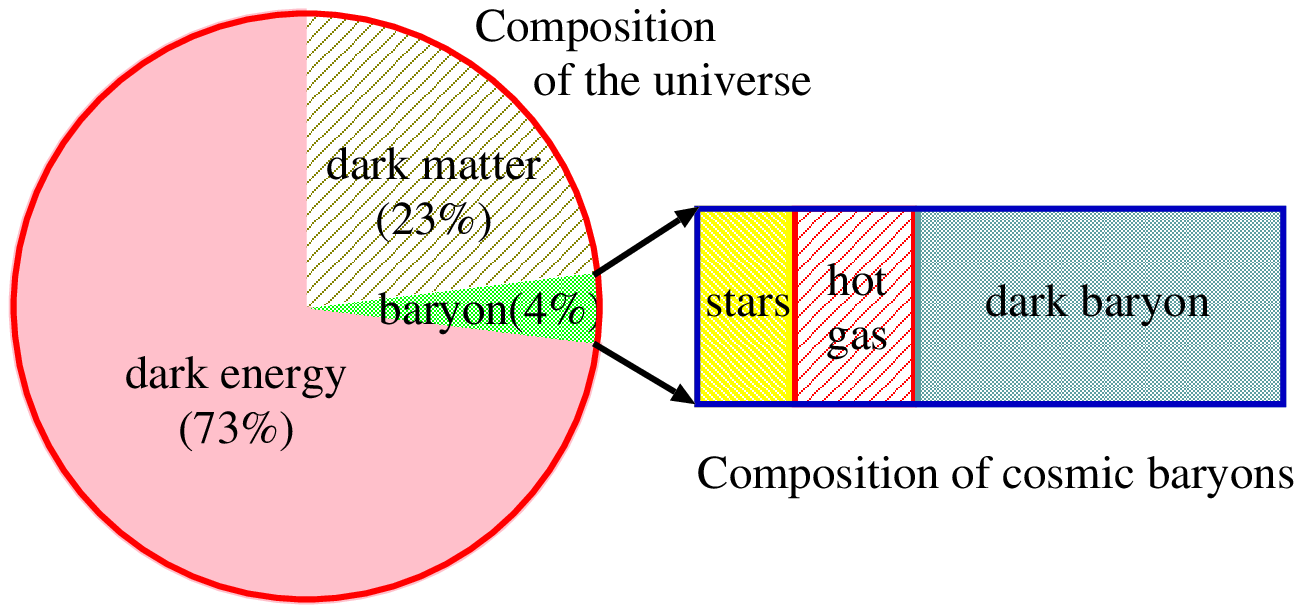}
\caption{Composition of the universe and cosmic baryons
\label{fig:composition}}
\end{center}
\begin{center}
\includegraphics[width=7cm]{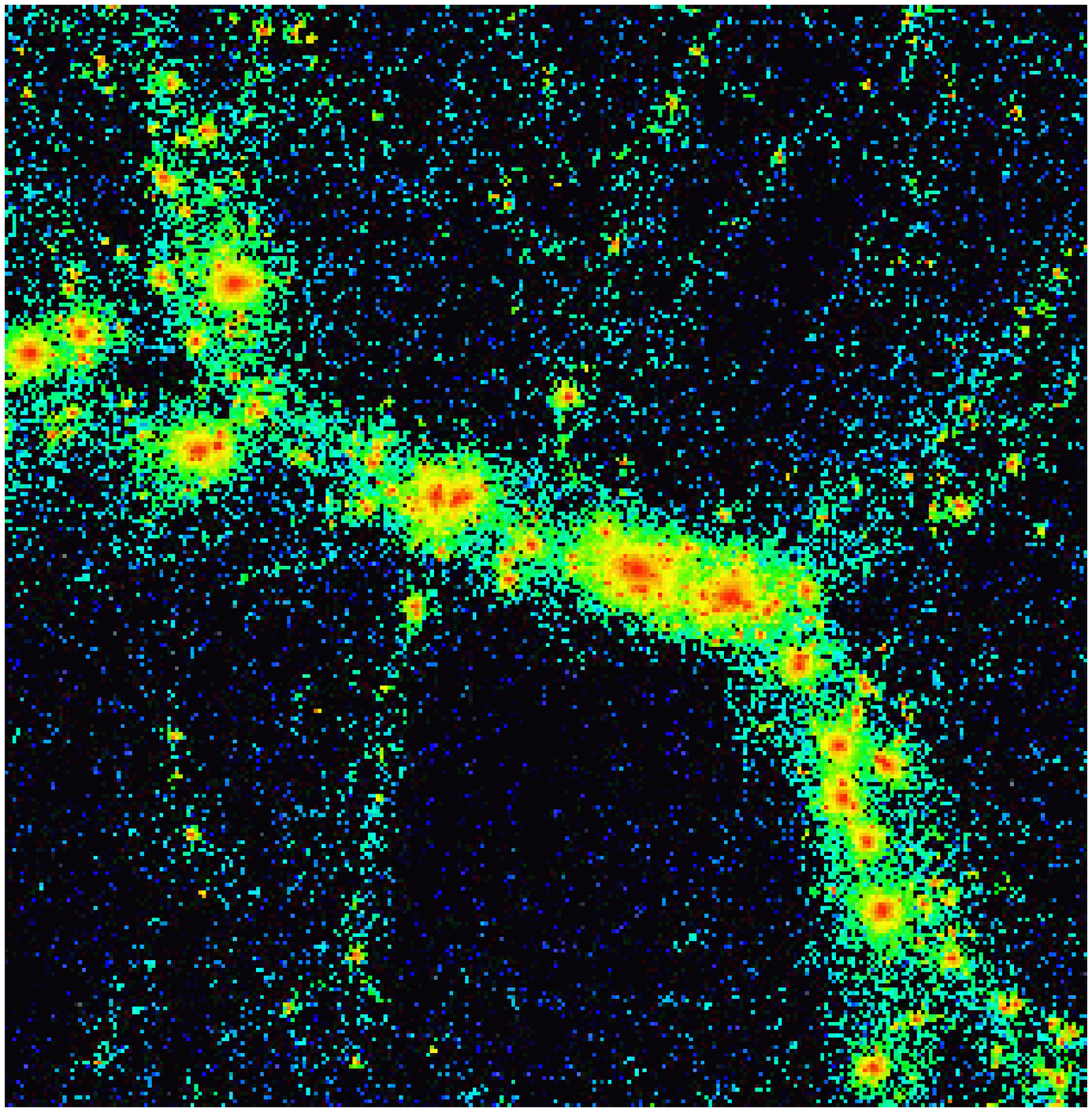}
\hspace*{0.2cm}
\includegraphics[width=7cm]{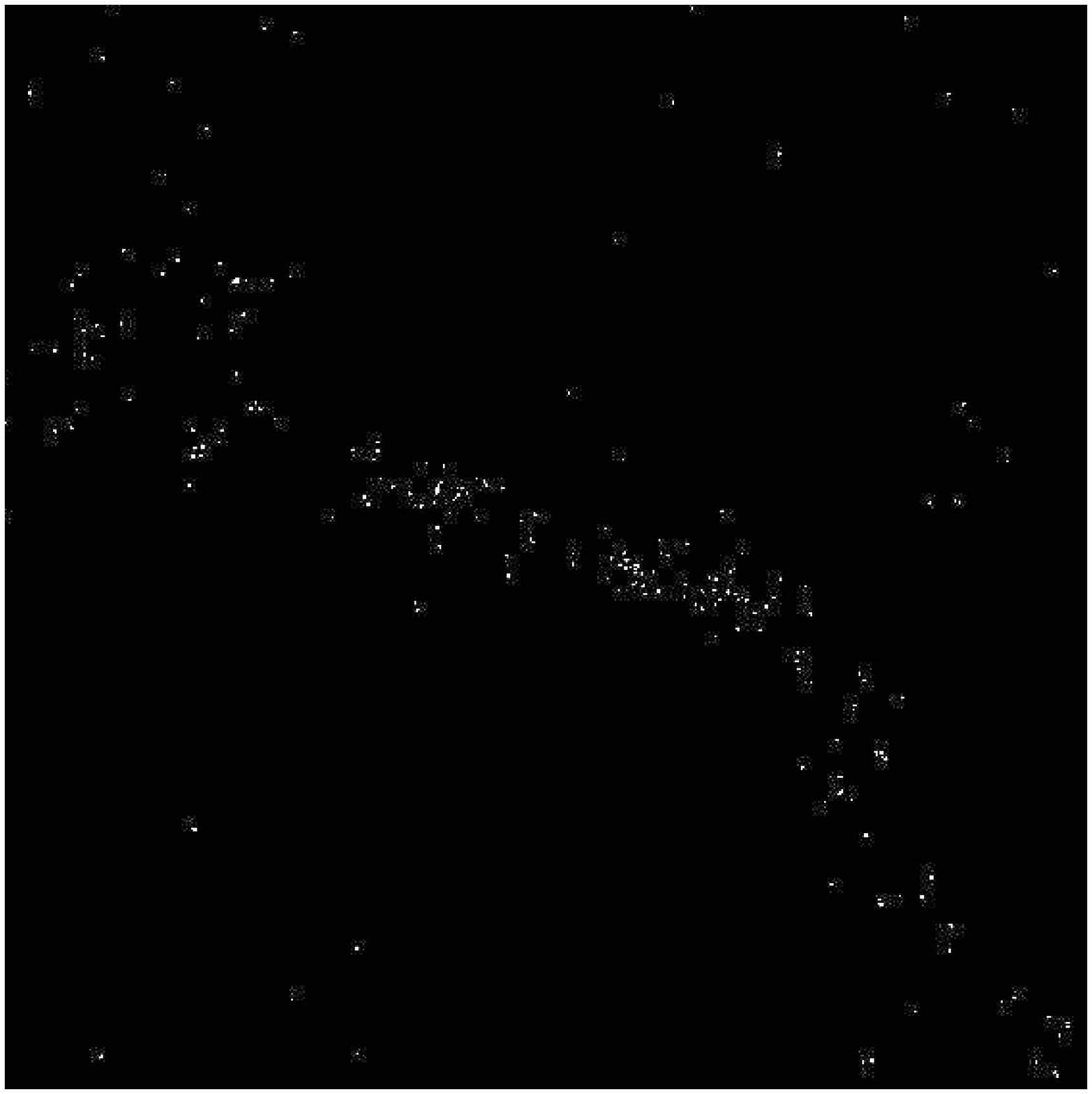}

\vspace*{0.3cm}

\includegraphics[width=7cm]{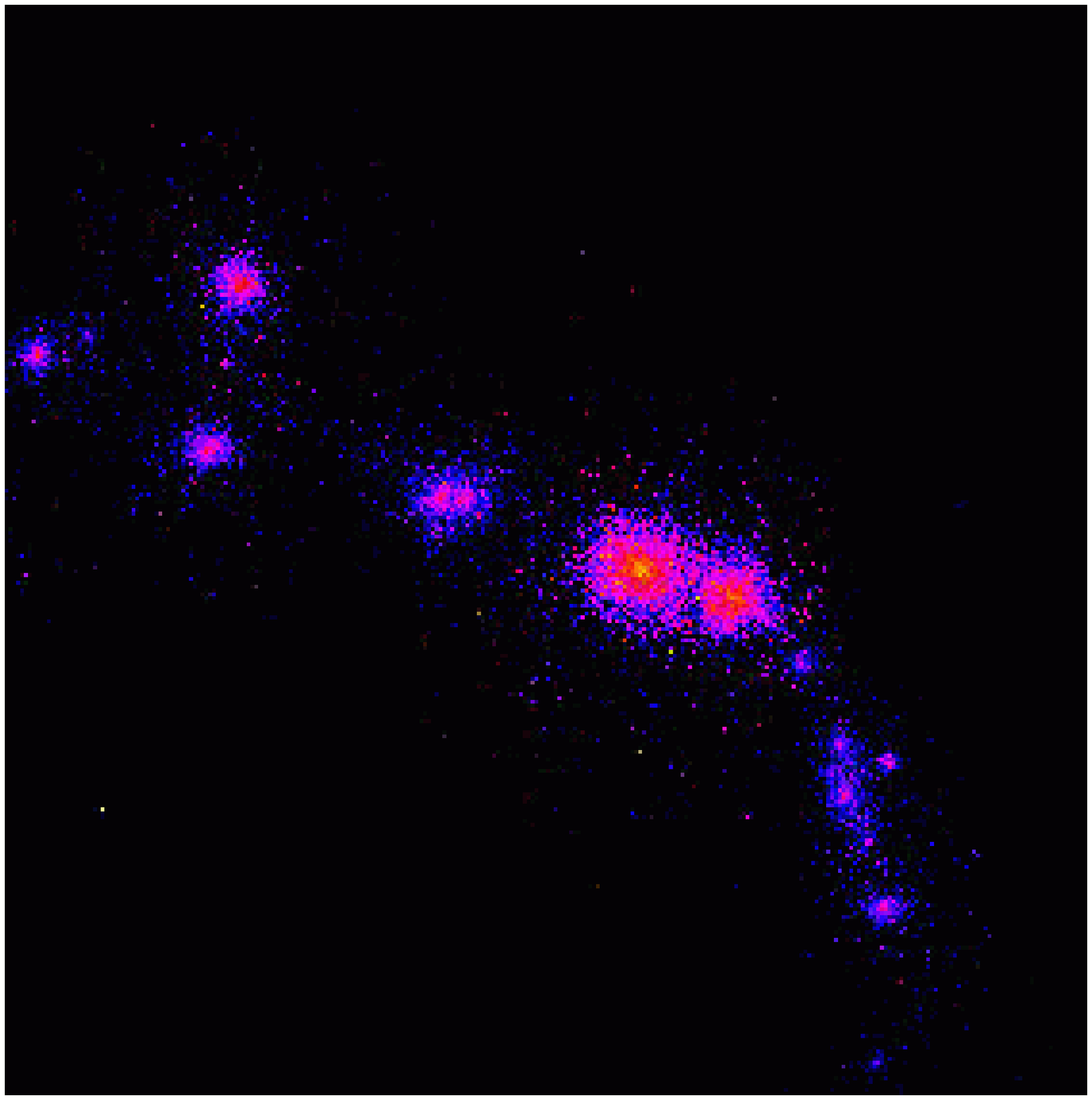} 
\hspace*{0.2cm}
\includegraphics[width=7cm]{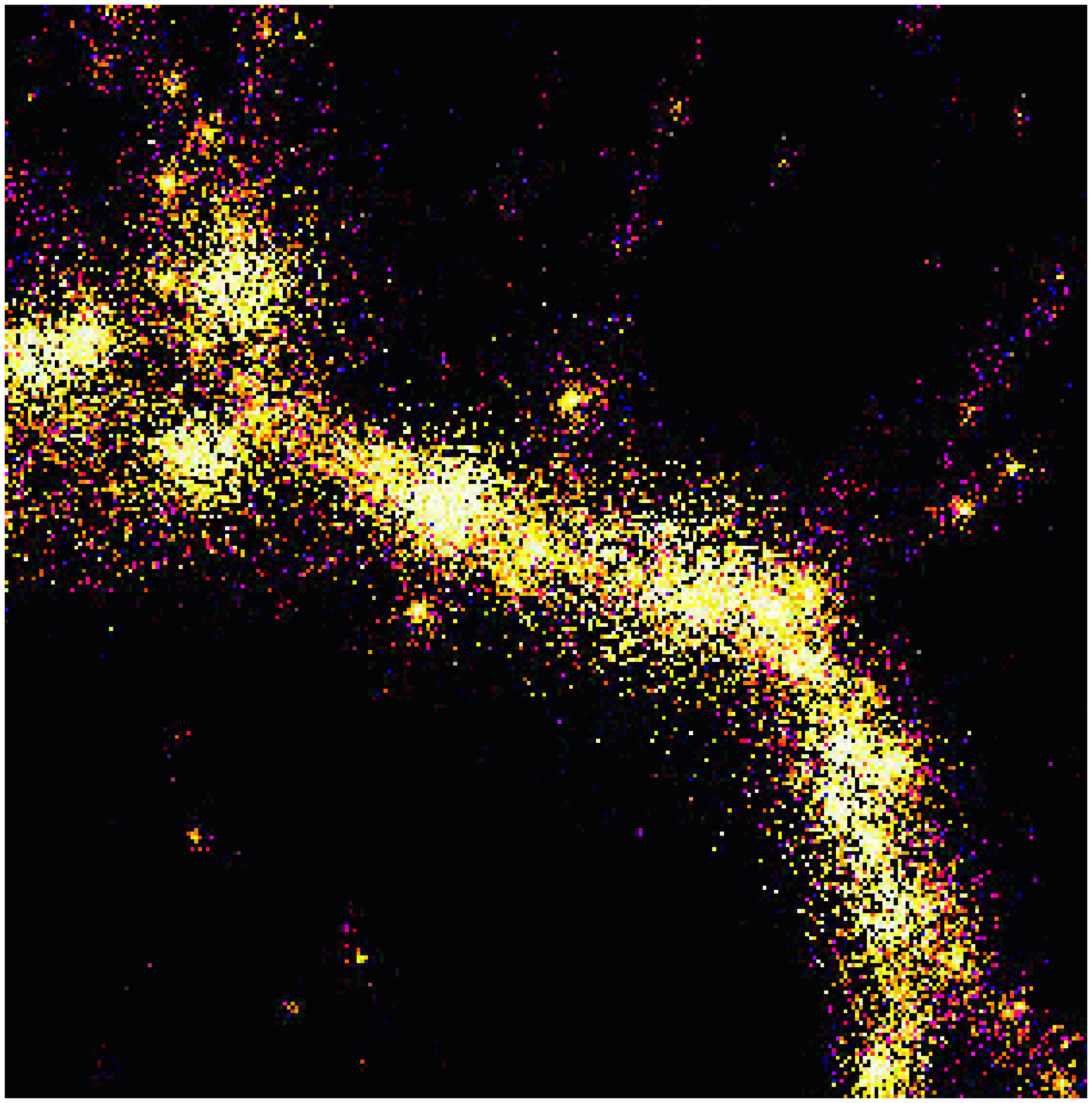} 
\caption{Simulated distribution of matter in the universe; {\it
Upper-left:} dark matter, {\it Upper-right:} galaxies (cold baryon
clumps below $T^4$K), {\it Lower-left:} hot intra-galactic medium ($T >
10^7 {\rm K}$), and {\it Lower-right:} warm-hot intergalactic medium
($10^5 {\rm K}< T < 10^7 {\rm K}$). The size of the plotted boxes
corresponds to $30h^{-1}$Mpc $\times 30h^{-1}$Mpc with the depth of
$10h^{-1}$Mpc.  \label{fig:distribution}}
\end{center}
\end{figure*}

It is widely accepted that our universe is dominated by {\it dark}
components; 23 percent in dark matter, and 73 percent in dark energy
\cite{Spergel03}.  More surprisingly, as Fukugita, Hogan \& Peebles
(1998) pointed out earlier\cite{Fukugita1998}, even the remaining 4
percent, cosmic baryons, has largely evaded the direct detection so far,
i.e., most of the baryons is indeed {\it dark} (see
Fig.\ref{fig:composition}). Those {\it cosmic missing baryons} may
consist of compact stellar objects (white dwarfs, neutron stars and
black holes), brown dwarfs, and/or diffuse gas.

 Subsequent numerical simulations\cite{Cen1999a,Cen1999b} indeed suggest
that approximately 30 to 50 percent of total baryons at $z=0$ take the
form of the warm-hot intergalactic medium (WHIM) with $10^5 {\rm K}< T <
10^7 {\rm K}$ which does not exhibit strong observational signature.
Figure \ref{fig:distribution} depicts snapshots of distribution of
different species of matter in the universe at $z=0$ from a smoothed
hydrodynamic simulation in a $\Lambda$ CDM universe
\cite{Yoshikawa2001};$\Omega_{\rm m}=0.3$, $\Omega_{\rm b}=0.015h^{-2}$,
$\Omega_{\Lambda}=0.7$, $\sigma_8=1.0$, and $h=0.7$, where $\Omega_{\rm
m}$ is the density parameter, $\Omega_{\rm b}$ is the baryon density
parameter, $\Omega_{\Lambda}$ is the dimensionless cosmological
constant, $\sigma_8$ is the rms density fluctuation top-hat smoothed
over a scale of $8h^{-1}\mbox{Mpc}$, and $h$ is the Hubble constant in
units of 100 km/s/Mpc. It employs $128^3$ dark matter particles and the
same number of gas particles in a comoving simulation cube of $L_{\rm
box}^3=(75h^{-1}\mbox{Mpc})^3$.  Clearly WHIM ($10^5 {\rm K}< T < 10^7
{\rm K}$; {\it lower right}) traces the large-scale filamentary
structure of mass (dark matter) distribution ({\it Upper left}) more
faithfully than hot intracluster gas ($T > 10^7 {\rm K}$; {\it Lower
left}) and galaxies ({\it Upper right}) both of which preferentially
resides in clusters that form around the knot-like intersections of the
filamentary regions. This implies that WHIM carries important
cosmological information in a complementary fashion to distribution of
galaxies (in optical) and of clusters (in X-ray).

Unfortunately the conventional X-ray emission of WHIM via the thermal
bremsstrahlung is very weak, and its detection has been proposed only
either through the O{\sc vi} absorption features in the QSO spectra or
the possible contribution to the cosmic X-ray background in the soft
band. Those attempts, however, are not well suited for unbiased
exploration of WHIM that is important for cosmological studies.  In
Ref.\cite{Yoshikawa2003}, we have proposed to survey WHIM using oxygen
emission lines.  

\clearpage

On the basis of this study, we propose a dedicated
soft-X-ray mission, {\it DIOS} (Diffuse Intergalactic Oxygen Surveyor),
which aims at performing a direct and homogeneous survey of WHIM that is
supposed to constitute the major fraction of cosmic missing baryons.

The detectability of oxygen emission lines with {\it DIOS} and the
current technical progress are discussed in Ref.\cite{Yoshikawa2003} and
Ref.\cite{Ohashi04}, respectively.

\section{DIOS}

We propose {\it DIOS}\, for a small satellite program which is a new
scheme under consideration by ISAS/JAXA. The primary scientific purpose
is a systematic sky survey of WHIM using oxygen K emission lines, O{\sc
vii} (561eV:1s$^2$--1s2s), O{\sc vii} (568eV:1s$^2$--1s2p), O{\sc vii}
(574eV:1s$^2$--1s2p), O{\sc vii} (665eV:1s$^2$--1s3p) and O{\sc viii}
(653eV:1s--2p). Thus both unprecedented energy resolution ($\Delta E
\approx 2$ eV) and large field of view are required.  These features
will be made possible by a combination of two innovations; a 4-stage
X-ray telescope and a large array of TES (Transition Edge Sensor)
micro-calorimeter.  {\it DIOS} will also perform a mapping observation
of the hot interstellar medium in our Galaxy.  Taking advantage of the
high energy-resolution, {\it DIOS} can detect the Doppler shifts of the
hot interstellar gas with a velocity $\sim 100$ km s$^{-1}$, directly
revealing the dynamics of heavy elements in hot bubbles in our Galaxy
(galactic fountain).

\begin{figure*}[thb]
\begin{center}
\includegraphics[width=16cm]{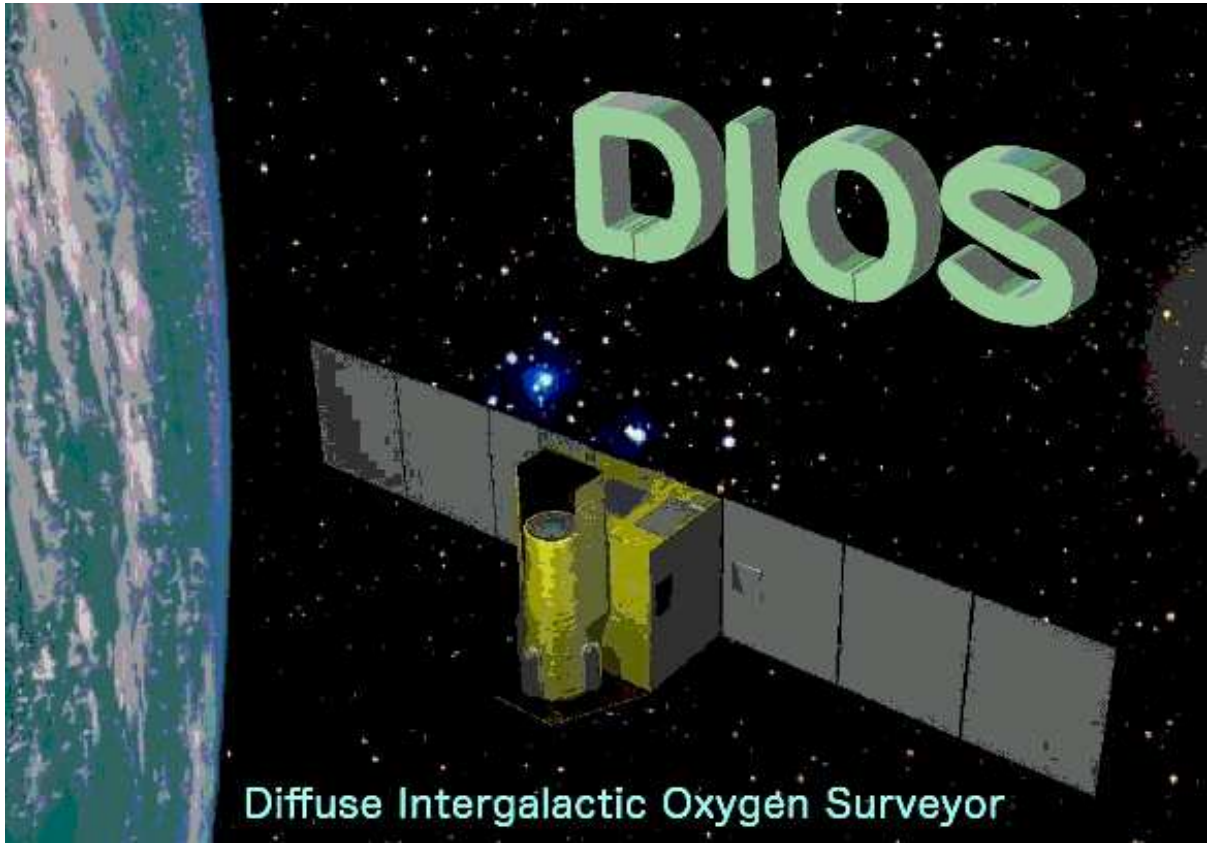} 
\caption{A dedicated soft X-ray mission to search for dark baryons 
via oxygen emission lines, {\it DIOS} (Diffuse Intergalactic Oxygen Surveyor).
The length of the solar paddle is 6 m. \label{fig:dios_spacecraft}}
\end{center}
\end{figure*}

\begin{table}[h]
\begin{center}
\caption{{\it DIOS} Mission summary  \label{tab:mission_summary}}
\begin{tabular}{ll}
\hline
Four-reflection X-ray telescope\\
\hline
FOV$\times$effective area & $S\Omega=$100 cm$^{2}$deg$^{2}$ at  0.6 keV\\
angular resolution  & $\sim 3$ arcmin \\
\hline
\hline
X-ray imaging spectrograph\\
\hline
energy range& $0.3 {\rm keV} <E< 1{\rm keV}$ \\
energy spectral resolution & 2eV　\\
size of detector & $>$10mm$\times$10mm \\
number of pixels  &　$\sim $16$\times$16  \\
FOV  &　$\sim 50'\times50'$  \\
\hline
\hline
Satellite system \\
\hline
orbit lifetime & $>1$ year\\
position control accuracy & $<0.5$ arcmin　\\
weight& $<400$ kg \\
\end{tabular}
\end{center}
\end{table}

\subsection{Spacecraft}

Figure \ref{fig:dios_spacecraft} shows a schematic view of the {\it
DIOS} spacecraft. Table \ref{tab:mission_summary} summarizes our
proposed features of {\it DIOS} for the successful WHIM detection (see
the next section).  The spacecraft will weigh about 400 kg in total
including the payload of $\sim 280$ kg. Thus the mission may be launched
also as a piggy-back or sub-payload in H2 or Ariane rockets. It is also
possible for a launch with the new ISAS rocket, such as M-V light. The
size before the launch is $1.5 \times 1.5 \times 1.2$ m, and the 1.2 m
side will be expanded to about 6 m after the paddle deployment.

The total power required is 500 W, of which 300 W is consumed by the
payload. The nominal orbit is a near earth circular one with an altitude
of 550 km. This low-earth orbit can be reached by the ISAS rocket M-V
light.  An alternative choice of the orbit under consideration is an
eccentric geostationary transfer orbit. This orbit gives a lower heat
input from the earth and relaxes the thermal design of the satellite,
and would enhance the launch opportunity to be carried as a sub-payload
for geostationary satellites. One significant drawback in this case,
however, is the increased particle background level as already
experienced by Chandra and XMM-Newton. In the soft energy range below 1
keV, electrons can be a major source of background.

The attitude will be 3-axis stabilized with momentum wheels. Typical
pointing accuracy will be about $10''$. The direction of the
field-of-view can be varied within $90^\circ \pm 20^\circ$ from the
Sun's direction. With this constraint, any position in the sky can be
accessed within half a year.

\subsection{Instruments}

\begin{figure}[h]
\begin{center}
\includegraphics[width=8.5cm]{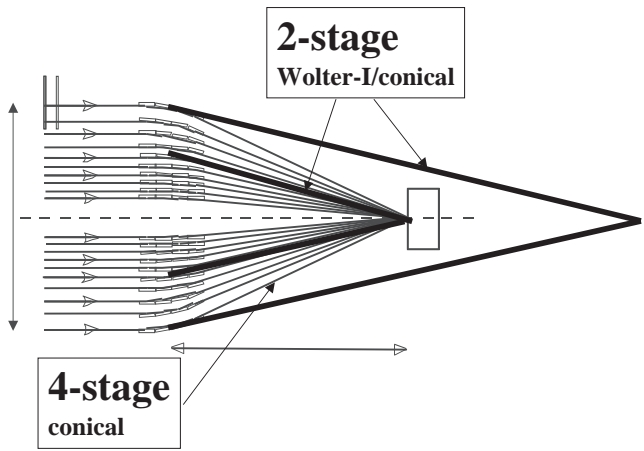}
\caption{Concept of the 4-stage reflection telescope.
\label{fig:FXT}}
\end{center}
\end{figure}

Several new technologies will be introduced in the {\it DIOS}
mission. The 4-stage X-ray telescope FXT (Four-stage X-ray Telescope) is
the first one \cite{tawara04}. As shown in Fig. \ref{fig:FXT}, incident
X-rays are reflected 4 times by thin-foil mirrors and are focused at
only 70 cm from the mirror level. At the energy of oxygen lines, $\sim
0.6$ keV, the reflectivity of the mirror surface is as high as 80\% and
the reduction of the effective area is not a serious problem. The
4-stage reflection makes the focal length around 50 percent of that in
the usual 2-stage design, substantially reducing the volume and weight
of the satellite. Also, a relatively small focal plane detector can have
a wide field of view, which is a great advantage for a TES calorimeter
array. In our basic design, the outer diameter of the mirror and the
effective area are 50 cm and 400 cm$^2$, respectively, at 0.6
keV. Ray-tracing simulation indicates that the angular resolution of is
2 arcmin (half-power diameter), and the image quality does not show
significant degradation at an offset angle $30'$. This 4-stage telescope
is a key design factor in making {\it DIOS} accommodated in the small
satellite package.

The focal plane instrument XSA (X-ray Spectrometer Array) is an array of
TES micro-calorimeters, whose development in Japan is our collaboration
with Waseda University, Seiko Instruments Inc., and Mitsubishi Heavy
Industries. There will be $16\times 16$ pixels covering an area of about
1 cm$^2$. The corresponding field of view is $50'$. XSA will have an
energy resolution of 2 eV (FWHM at 0.6 keV).  We are now developing
several new techniques toward the multi-pixel operation of TES
calorimeters. X-ray absorbing material under consideration is Bi, which
has low heat capacity and does not produce long-living quasi particles
in the absorber. We are testing an electro-plating method to build a $16
\times 16$ array with a pixel size $\sim 0.5 \times 0.5$ mm supported by
a thin stem. For the signal readout, efficient multiplexing of the
signals is essential to take out all the data from the cold stage. We
are trying to add the signals in frequency space by operating the TES
calorimeters with AC bias at different frequencies. We were successful
in decoding signals from 2 different pixels so far.  A new multi-input
SQUID has also been developed to add the signals from several TES pixels
together. An efficient thermal shield with high soft X-ray transmission
is an essential item, and we are considering very thin Be foils for this
purpose.

Another important feature of {\it DIOS} is the cryogen-free cooling
system. We are considering a serial connection of different types of
coolers to achieve $\sim 50$ mK for XSA within the available power
budget. In the first stage, a Stirling cooler takes the temperature down
to 20 K, and then $^3$He Joule-Thomson cooler reduces it to 1.8 K as the
second stage. The third stage is $^3$He sorption cooler achieving
0.4 K, and finally adiabatic demagnetization refrigerator obtains the
operating temperature at 50 mK. Since no cryogen is involved, this
cooling system ensures an unlimited observing life in the orbit, which
is a big advantage of {\it DIOS}. The XSA system is subject to a warm
launch, therefore we have to allow for the initial cooling of the system
for the first 3 months in the orbit.

Since several challenging technologies are involved, we seriously
consider international collaboration in various parts of the observing
instruments.

\begin{figure*}[thb]
\begin{center}
\includegraphics[width=16cm]{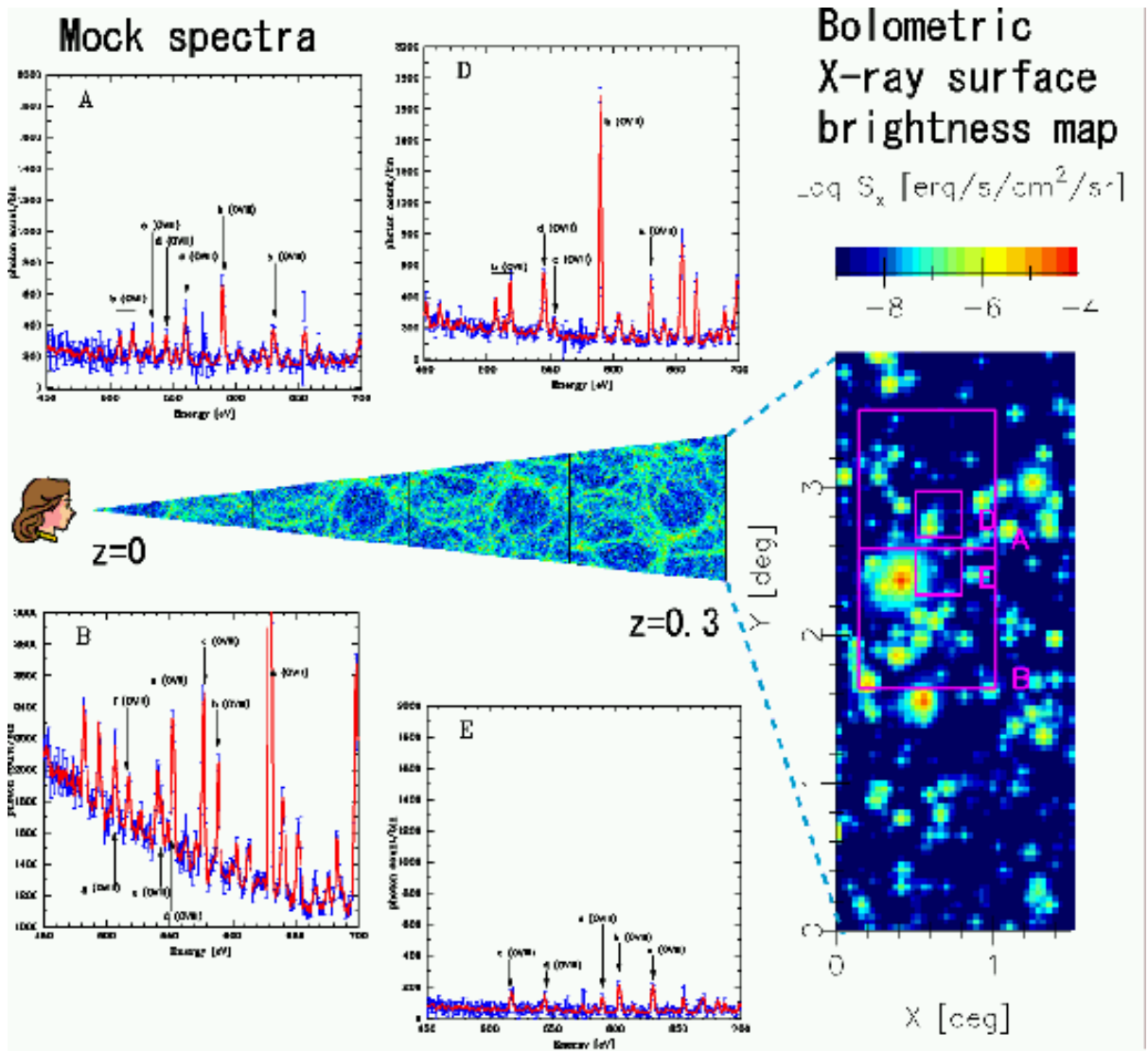} 
\caption{Mock surface density maps for bolometric X-ray emissions
and the corresponding high energy-resolution soft-X-ray 
spectra of WHIMs expected for {\it DIOS}.
\label{fig:mockobs}}
\end{center}
\end{figure*}

\section{Detectability of Warm/Hot Intergalactic Medium 
with DIOS}

\subsection{Constructing mock spectra from cosmological 
hydrodynamic simulations}

We have examined the detectability of WHIM through O{\sc viii} and O{\sc
vii} emission lines with {\it DIOS} in detail previously
\cite{Yoshikawa2003} assuming a detector which has a large throughput
$S_{\rm eff}\Omega=10^2$ cm$^2$ deg$^2$ and a high energy resolution
$\Delta E = 2$ eV.  Here we briefly summarize those results.

\begin{figure}[thb]
 \begin{center}
\includegraphics[width=8.5cm]{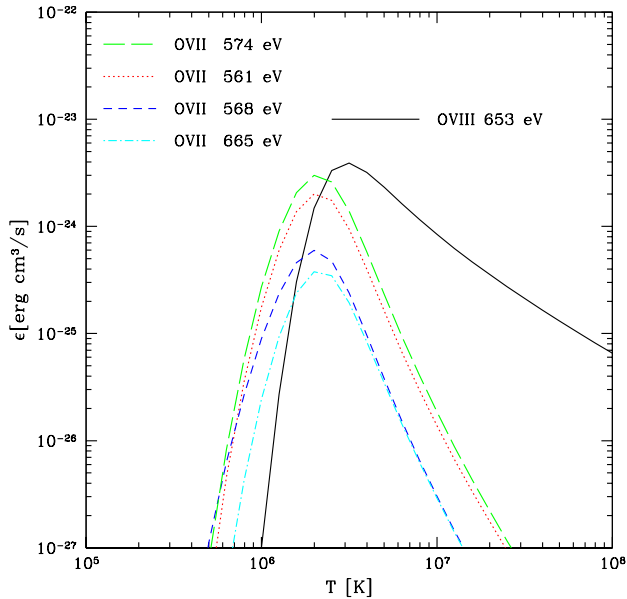}
  \caption{Emissivity of O{\sc vii} and O{\sc viii} lines in collisional
  ionization equilibrium. \label{fig:emissivity}}
 \end{center}
\end{figure}

First we create the simulation lightcone data over the $5^\circ \times
5^\circ$ region of a sky up to $z=0.3$ (see Fig.\ref{fig:mockobs}); we
put $64\times 64$ square grids on the celestial plane and 128
equally-spaced bins along the redshift direction.  We compute the
surface brightness of each cell on the $64\times64\times128$ grid, and
integrate along the same line of sight.  The emissivities of O{\sc vii}
and O{\sc viii} lines (Fig.~\ref{fig:emissivity}) are computed assuming
the collisional ionization equilibrium.  For simplicity, the metallicity
is set to be $Z=0.2Z_{\odot}$ independently of redshifts and the
densities of the intergalactic medium.  This procedure yields the
projected surface density map ($64\times64$ pixels) on the celestial
plane.

\begin{figure}[thb]
 \begin{center}
\includegraphics[width=8.5cm]{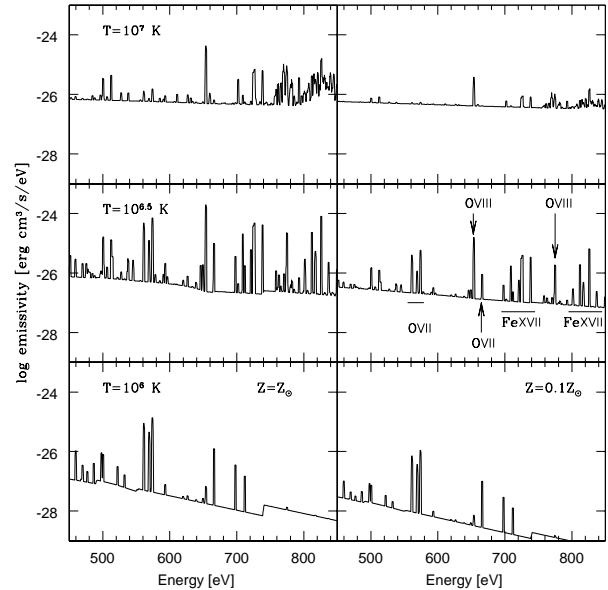}
 \caption{Template spectra of collisionally ionized plasma with
 temperature $T=10^6$ K ({\it lower panels}), $10^{6.5}$ K ({\it middle
 panels}), and $10^7$ K ({\it upper panels}). Spectra for metallicity
 $Z=Z_{\odot}$ and $Z=0.1Z_{\odot}$ are shown in the left and right
 panels, respectively. \label{fig:template}}
 \end{center}
\end{figure}

Similarly we construct the corresponding mock spectra at soft X-ray
energy band ranging from 450eV to 700eV along each line of sight.
Figure~\ref{fig:template} shows examples of the template spectra that we
adopt.  At a lower temperature, $T=10^6$ K, we have strong emission
lines of the O{\sc vii} triplets ($E=561, 568, 574$ eV). Around
$T=10^{6.5}$ K and $10^{7}$ K, O{\sc viii} line at $E=653$ eV and many
Fe\,{\sc xvii} lines at $E>700$ eV become prominent.

\begin{figure}[htb]
 \begin{center}
\includegraphics[width=8.5cm]{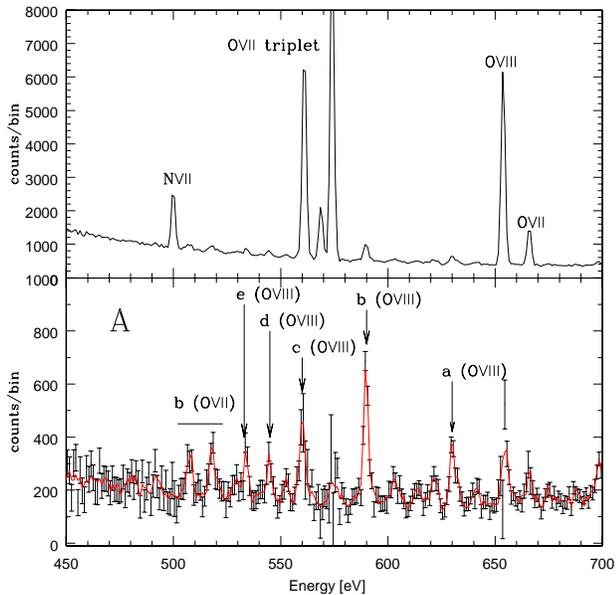}
 \caption{A simulated spectrum along a line of sight. {\it Upper panel}
 shows emission lines of WHIM, the CXB and the Galactic emission. {\it
 Lower panel}: The spectrum of WHIM after the CXB and the Galactic
 emission are subtracted. The constant metallicity of $Z=0.2Z_\odot$ is
 adopted.\label{fig:spec_subtracted}}
 \end{center}
\end{figure}

The upper panel of Figure~\ref{fig:spec_subtracted} shows an example of
the simulated composite spectrum, which includes the contribution from
WHIM, Cosmic X-ray Background, the Galactic emission lines, with
exposure time $T_{\rm exp}=3\times10^5$ sec for the region A (0.88
deg$^2$) in Fig.~\ref{fig:mockobs}.  Strong lines in the upper panel
correspond to the Galactic emission lines of N{\sc vii} at 498 eV, O{\sc
vii} at 561eV, 568 eV, 574 eV, and 665 eV, and O{\sc viii} at 653 eV.
Clearly the separation of the Galactic component from the WHIM emission
lines is the most essential.  In order to mimic the realistic separation
procedure, we construct an independent realization of spectra which
consists purely of the CXB and Galactic emission lines but using the
different sets of random numbers in adding the Poisson fluctuations in
each bin.  Then the latter spectra are subtracted from the mock spectra
(WHIM + CXB and Galactic emissions), which yields a residual mock
spectrum purely for the WHIM ({\it dots with Poisson error bars in lower
panel}).  For comparison, we plot the noiseless WHIM spectrum from
simulation in solid line.  The emission lines with labels in the lower
panel indicate O{\sc viii} and O{\sc vii} triplet lines whose surface
brightness exceeds the expected limiting flux of {\it DIOS}.  Thus this plot
indicates that the emission lines exceeding the residual photon number
$\sim 100$ counts/bin are detectable with a three-day exposure.

\subsection{Results}

Figure \ref{fig:mockobs} summarizes the mock imaging map and the
corresponding spectra for the {\it DIOS} survey.  Since WHIM is supposed
to reside in the outskirts of galaxy clusters and/or in galaxy groups,
it is quite natural to select vicinity of rich galaxy clusters as target
regions.  Thus we select the regions A and B for shallower exposure
($T_{\rm exp}=3\times10^5$ sec, field-of-view of 0.88 deg$^2$, and
$S_{\rm eff}\Omega T_{\rm exp}=3\times10^7$ cm$^2$ deg$^2$ sec) and D
and E for deeper exposure ($T_{\rm exp}=10^6$ sec, field-of-view of
0.098 deg$^2$, and $S_{\rm eff}\Omega T_{\rm exp}=1.1\times10^7$ cm$^2$
deg$^2$ sec).

The region B encloses an X-ray cluster located at $z=0.038$, and the
region A contains a filamentary structure around the cluster
(Fig.~\ref{fig:mockobs}). The spectra along the region B exhibit a
strong O{\sc viii} emission line at $E=629$ eV, which originates from
the intra-cluster medium of the X-ray cluster at $z=0.038$. The O{\sc
vii} triplet emission lines around $E=535-560$ eV are also ascribed to
the same cluster.  The region D shows an emission line due to a
substructure of a cluster at $z=0.039$.  The spectra in the regions D
and E show the emission lines from the galaxy cluster and its
substructure at $z=0.039$.  We note that the emission line in the region
D corresponding to the $z=0.039$ structure is stronger than the
counterpart in the region E, although the temperature of the region D at
$z=0.039$ is $\simeq 7\times10^6$ K and in fact lower than that of the
region E ($\simeq 2\times10^7$ K).  This is because the emissivity of
O{\sc viii} decreases as the temperature exceeds $T \approx 3\times10^6$
K, and clearly demonstrates that the oxygen lines are more sensitive to
the presence of the WHIM than that of the higher temperature gas
associated with intra-cluster gas.

Figure~\ref{fig:mockobs} indeed demonstrates that the high-spectral
resolution of {\it DIOS} enables to identify the WHIM at different
emission energies, i.e., Oxygen emission line tomography of the WHIMs at
different locations.

\subsection{Observational strategy}

\begin{figure}[thb] 
\begin{center}
\includegraphics[width=6.5cm,angle=270]{1000photons.eps}
\caption{Pulse-height spectrum for OVII and OVIII lines expected with
{\it DIOS} from a plasma with $kT = 0.2$ keV when 1000 line photons are
accumulated.
\label{fig:1000photons}}
\includegraphics[width=8.5cm]{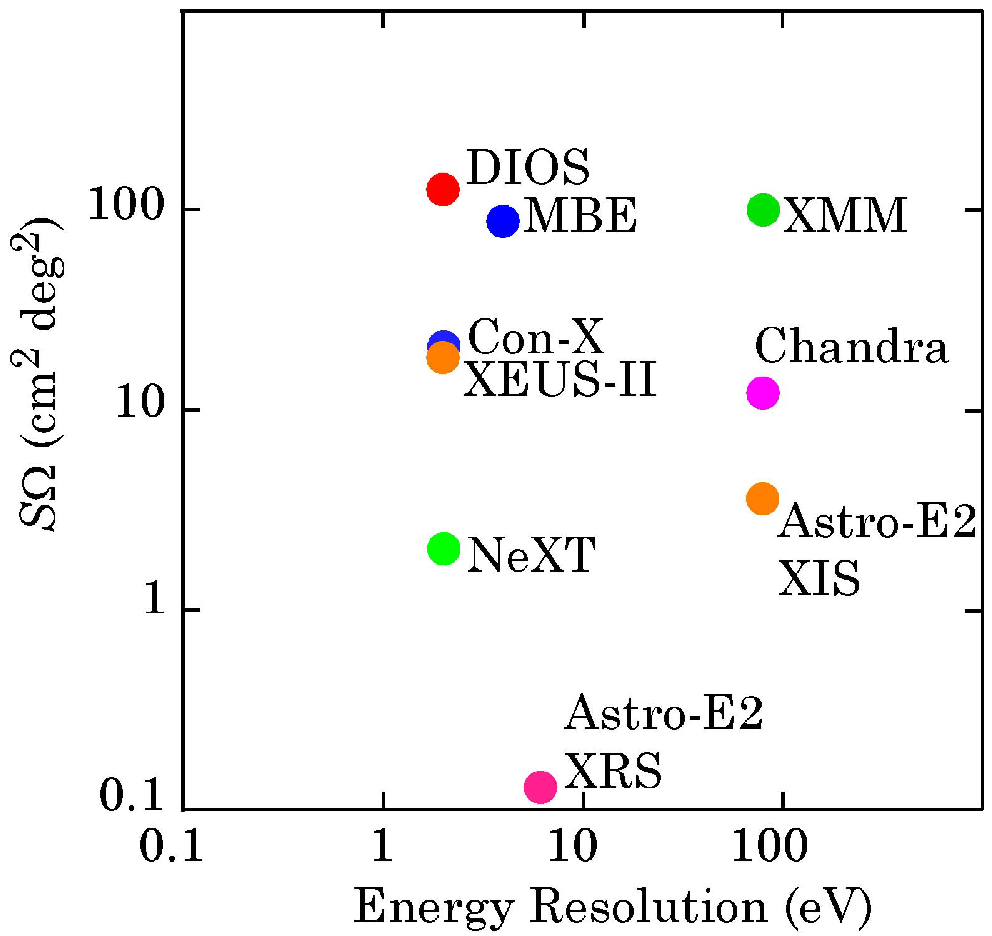}
\caption{Comparison of $S\Omega$ and energy resolution for
spectroscopic instruments (CCDs and micro-calorimeters) for planned and
operating X-ray satellites.
\label{fig:hikaku}}
\end{center}
\end{figure}

The previous subsection indicates that a single pointing needs typically
100 ksec to obtain enough oxygen photons from WHIM. Figure
\ref{fig:1000photons} shows the expected pulse-height spectrum of 0.2keV
plasma when 1000 photons of oxygen emission lines are collected. We
expect that this quality of data can be obtained from surrounding
regions of clusters of galaxies with 100--300 ksec of observation. For
large-scale filaments, roughly 10 times less photos are expected for the
same observing time. On the other hand, the hot interstellar medium in
our galaxy produces an order-of-magnitude stronger line emission. With
this data quality, the lines in the OVII triplet are clearly detected
and we will be able to measure the temperature of the WHIM
directly. These three lines can also be used to separate individual
plasma components when several emission regions with different redshifts
overlap along the line of sight.

Effective observing time will be approximately 40 ksec for a near earth
orbit. So if one integrates for 100 ksec in each pointing position, a
sky map for an area of $10^\circ \times 10^\circ$ can be produced in a
year. This size of the survey area is sufficient to trace the
large-scale structure of the universe at $z < 0.3$. This survey
observation of a limited sky will be the first task of {\it DIOS}. Since
oxygen lines from the galactic interstellar medium are stronger by
roughly 2 orders of magnitude, 1 ksec in each point is good to make a
large map of the interstellar medium distribution. We plan to devote the
second year for a survey of a $100^\circ\times 100^\circ$ sky for the
Galactic interstellar medium. After these 2 years, further deeper
observations of the intergalactic medium as well as of the outflowing
hot gas in near-by galaxies can be performed.

Figure \ref{fig:hikaku} compares $S\Omega$ and energy resolution among
different instruments in the X-ray missions which are planned or already
in operation. Clearly, {\it DIOS} will achieve the highest sensitivity
for soft X-ray lines from extended objects. We believe that {\it DIOS},
with its complementary performance to larger general-purpose X-ray
missions, will bring very rich science on cosmological evolution of
baryons.

\section{Conclusions}

We have presented our recent proposal of a dedicated soft-X-ray mission,
{\it DIOS} (Diffuse Intergalactic Oxygen Surveyor).  Within the exposure
time of $T_{\rm exp}=10^{5-6}$ sec {\it DIOS} will be able to reliably
identify O{\sc viii} emission lines (653eV) of WHIM with $T=10^{6-7}$ K
and the overdensity of $\delta=10^{0.5-2}$, and O{\sc vii} emission
lines (561, 568, 574, 665eV) of WHIM with $T=10^{6.5-7}$ K and
$\delta=10^{1-2}$.  The WHIM in these temperature and density ranges
cannot be detected with the current X-ray observations except for the
oxygen absorption features toward bright QSOs. {\it DIOS} is especially
sensitive to the WHIM with gas temperature $T=10^{6-7}$K and overdensity
$\delta=10-100$ up to a redshift of $0.3$ without being significantly
contaminated by the cosmic X-ray background and the Galactic emissions.
There is a similar proposal called Missing Baryon Explorer (MBE) in the
U.S. \cite{Fang2004}. {\it DIOS}, hopefully launched in several years,
promises to provide yet another important and complementary tool to
trace the large-scale structure of the universe via dark baryons.

\begin{acknowledgments}
This research was supported in part by the Grant-in-Aid for Scientific
Research of Japan Society of Promotion of Science.  Numerical
computations presented in this paper were carried out mainly at ADAC
(the Astronomical Data Analysis Center) of the National Astronomical
Observatory, Japan (project ID: yys08a, mky05a).
\end{acknowledgments}

\end{document}